\documentclass[prl,aps,twocolumn,floats,nofootinbib,showpacs]{revtex4}

\usepackage{amsmath,amssymb,dsfont,graphicx,psfrag}
\usepackage{epsfig}

\begin{document}
\title{Superconductor-Insulator Transitions and Magnetoresistance Oscillations in Superconducting Strips}

\author{Yeshayahu Atzmon$^{1}$ and Efrat Shimshoni$^{1}$ }
\affiliation{$^{1}$Department of Physics, Bar-Ilan University, Ramat-Gan 52900, Israel }

\date{\today}
\begin{abstract}
The magnetoresistance of thin superconducting (SC) strips subject to a perpendicular magnetic field $B$ and low temperatures $T$ manifests a sequence of alternating SC--insulator transitions (SIT). We study this phenomenon within a quasi one-dimensional (1D) model for the quantum dynamics of  vortices in a line-junction between coupled parallel SC wires, at parameters close to their SIT. Mapping the vortex system to 1D Fermions at a chemical potential dictated by $B$, we find that a quantum phase transition of the Ising type occurs at critical values of the vortex filling, from a SC phase near integer filling to an insulator near $1/2$--filling. For $T\rightarrow 0$, the resulting magnetoresistance $R(B)$ exhibits oscillations similar to the experimental observation.

\end{abstract}

\maketitle

The conduction properties of low--dimensional superconducting (SC) systems (thin films and wires) are strongly dominated by fluctuations in the SC order parameter. A particularly prominent manifestation of the role of fluctuations is the appearance of a finite dissipative resistance below the mean--field critical temperature $T_c$ of the bulk superconductor. At low temperatures $T\ll T_c$, the dominant fluctuations are in the {\it phase} of the complex order parameter. Most notably, topological excitations (vortices and phase--slips) can generate dissipation in their liquid state.
In the $T\rightarrow 0$ limit, their quantum dynamics becomes significant and may drive a transition to a metallic or insulating state \cite{SITrev,SIT1D}.

In the one--dimensional (1D) case, i.e. SC wires of width and thickness smaller than the coherence length $\xi$, the resistance essentially never vanishes at finite $T$ due to thermal activation of phase--slips \cite{LAMH,TAPS} (for $T\lesssim T_c$) or their quantum tunneling at lower $T$ \cite{Giordano,zaikin,SIT1D}. In contrast, in the 2D case (SC films) superconductivity is well-established at sufficiently low $T$.
However, a quantum ($T\rightarrow 0$) superconductor--insulator transition (SIT) \cite{QPT2,SITrev}
can be tuned by an external parameter which leads to proliferation of free vortices. Employing charge--flux duality \cite{Fisher} it is possible to view the SC phase as a vortex solid, and the insulator as a vortex superfluid.

A convenient means of inducing a SIT in SC films is by application of a perpendicular magnetic field $B$. At fixed $T$, a positive magnetoresistance $R(B)$ is typically observed in a wide range of $B$. The SIT is then clearly indicated in the data as a crossing point of these isotherms at a critical field $B_c$, separating a SC phase (where $dR/dT>0$) for $B<B_c$ from the insulating phase ($dR/dT<0$) for $B>B_c$.

\begin{figure}
\includegraphics[width=1.0\linewidth]{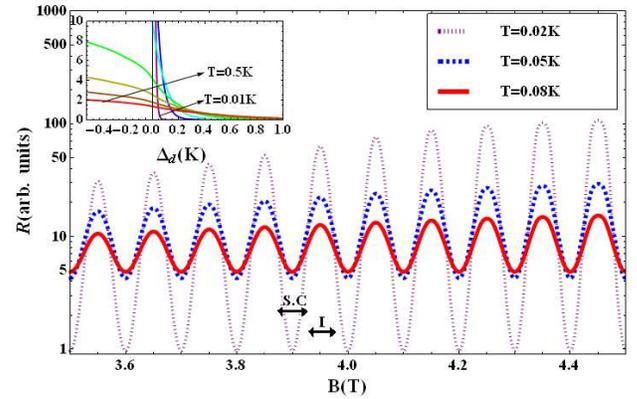}
\caption{(color online)
Isotherms of $R$ as a function of
$B$, for $J=1.06$K, $V=0.98$K, $v_-/L=0.01$K [see Eq. (\ref{H_-_ferm})], indicating superconducting (SC) and insulating (I) sections alternately. Inset: $R$ vs. the gap $\Delta_d$ [see Eq. (\ref{gaps}) and text therafter] near a single critical
point, for $T=0.01,0.05,0.1,0.2,0.3,0.4,0.5$K.
  \label{fig1} }
\end{figure}

A recent experimental study of a strip geometry \cite{shahar} -- namely, a SC wire of width comparable to $\xi$ -- offers an opportunity to probe the crossover from a 1D to 2D quantum dynamics of the topological phase--defects in SC devices. The prominent observation is that in the presence of a perpendicular field $B$, the magnetoresistance $R(B)$ exhibits oscillations which amplitude is sharply increasing at low $T$, in striking resemblance to the behavior of Josephson arrays \cite{glazman} and SC network systems \cite{networks}. Moreover, the SIT at $B_c$ appears to be preempted by several consecutive transitions at lower fields, from a SC to an insulator or vice versa alternately.

The periodicity of the above mentioned oscillations is consistent with a single flux penetration to the sample, suggesting that the observed SC or insulating behavior of the system is determined by commensuration of vortices within the strip area. When an integer number of vortices can be fitted along the strip length, superconductivity may be supported even at sufficiently high $B$ such that a large fraction of the sample area turns normal. Deviation from commensurability of the vortex filling weakens superconductivity, possibly inducing a transition to a metallic \cite{glazman} or insulating state.

In this paper we focus on the strongly quantum fluctuation regime characterizing the 1D vortex matter close to an SIT, and propose a
theory for its low $T$ transport behavior. The system is shown to
exhibits a series of quantum phase transitions of the Ising type, manifested as SC--insulator
oscillations of the Ohmic resistance $R(T,B)$ [see Fig. \ref{fig1}]. This result underlines a correspondence between charge-flux duality across a SIT and the order-disorder duality characteristic of the Ising transition.

We consider a SC strip subject to a perpendicular magnetic field $B\lesssim B_c$.
Assuming that the high vortex
density in this case leads to near merging of their cores along the central axis of the strip, we model the system as
a line--junction formed by a pair of parallel
SC wires of length $L\gg \xi$, separated by a thin normal barrier of width $w$.
In the low $T$ (phase-fluctuations) regime, the dynamics of the collective
phase field in the wires ($\phi_i(x,t)$ with $i=1,2$) is governed by the
effective 1D Hamiltonian
\begin{equation}\label{Hfull}
H_0=H_1+H_2+H_{int}\; ,
\end{equation}
in which (using units where $\hbar=1$)
\begin{eqnarray}\label{H_i}
H_i&=&\frac{1}{2}\int_{-\frac{L}{2}}^{\frac{L}{2}} dx
\left[U_0\rho_i^2 +
\frac{\rho_s}{4m} (\partial_x \phi_i)^2 \right]\; , \\
H_{int}&=&\int_{-\frac{L}{2}}^{\frac{L}{2}} dx
\left[-g_J\cos(\phi_1-\phi_2-qx) +U\rho_1\rho_2\right].
\label{Hint}
\end{eqnarray}
Here the operator $\rho_i(x)$ denotes density fluctuations of
Cooper pairs in wire $i$, and can be represented as \cite{book}
\begin{equation}\label{rhototheta}
\rho_i(x)=-\frac{1}{\pi}\partial_x
\theta_i(x)+\rho_0\sum_{p\not=0}e^{i2p(\pi\rho_0x-\theta_i(x))}
\end{equation}
in terms of the conjugate field $\theta_i(x)$ satisfying
$[\phi_i(x),\partial_x\theta_i(x^\prime)]=i\pi\delta(x^\prime-x)$.
The first term in Eq. (\ref{H_i}) hence describes a charging
energy; $\rho_s$ is the superfluid density (per unit length) assumed to be monotonically suppressed by increasing $B$, $\rho_0=\rho_s(B=0)$ and $m$ is the electron mass. The inter--wire coupling [Eq.
(\ref{Hint})] consists of a Josephson term and an
inter--wire Coulomb interaction, of coupling strengths $g_J$ and
$U$, respectively; $q=2\pi wB/\Phi_0$ (with $\Phi_0$ the flux
quantum) denotes vortex density per unit length. $H_0$
describes an ideal system, to which we later add a disorder
scattering potential.

It is convenient to introduce total and relative phase fields via
the canonical transformation $\phi_\pm=(\phi_1\pm\phi_2)/\sqrt{2}$
and the corresponding conjugate fields
$\frac{1}{\pi}\partial_x\theta_\pm$, in terms of which the
Hamiltonian (\ref{Hfull}) is separable:
\begin{equation}
\label{Hpm}
 H_0=H_++H_-
\end{equation}
\begin{multline}
{\rm where}\quad H_+=H_{LL}^{(+)}\; , \quad
H_-=H_{LL}^{(-)}+  \\ +\int_{-\frac{L}{2}}^{\frac{L}{2}} dx\left[-g_J\cos(\sqrt{2}\phi_--qx)
+ g_c\cos(\sqrt{8}\theta_-)\right] \; ; \\
 H_{LL}^{(\pm)}\equiv
\frac{v_\pm}{2\pi}\int_{-\frac{L}{2}}^{\frac{L}{2}} dx \left[ K_\pm
(\partial_x\theta_\pm)^+\frac{1}{K_\pm} (\partial_x \phi_\pm)^2
\right]\label{HLL}
\end{multline}
and the parameters are given by
\begin{eqnarray}
\quad K_\pm &=&\sqrt{\frac{4m(U_0\pm U)}{\pi^2\rho_s}}\; ,\quad
v_\pm=\sqrt{\frac{\rho_s(U_0\pm U)}{4 m}}\; , \nonumber\\
g_c&=&2U\rho_0^2\; . \label{Kvg_c}
\end{eqnarray}
Here we have accounted for the most relevant interaction terms,
neglecting umklapp terms that are effectively suppressed due to
the rapidly oscillating factor in Eq. (\ref{rhototheta}).

We next define new canonical fields
\begin{equation}
\phi  \equiv \frac{1}{{\sqrt 2 }}\phi _ -\; ,\quad \theta  \equiv \sqrt{2} \theta _ -
\label{new_theta}
\end{equation}
in terms of which $H_{LL}^{(-)}$ acquires the form of a Luttinger Hamiltonian with an effective Luttinger parameter $K = K_-/2$.
Assuming that the original parameter $K_-$ is close to the critical value for a SIT in a 1D wire, $K_c=2$ \cite{zaikin}, we obtained $K\approx 1$. This yields
\begin{eqnarray}
H_- &=& \frac{v_-}{2\pi}\int_{-\frac{L}{2}}^{\frac{L}{2}}  \left[ (\partial _x  \theta )^2  + (\partial _x  \phi )^2  \right] \\
 &+& \int_{-\frac{L}{2}}^{\frac{L}{2}}dx \left[-g_J\cos (2 \phi -qx ) + g_c\cos (2 \theta ) \right]\; .\nonumber
\label{H_-bos}
\end{eqnarray}

The model Eq. (\ref{H_-bos}) can be refermionized by introducing right ($R$) and left ($L$) moving Fermion fields \cite{SNT}
\begin{equation}
\psi _{R,L} = \frac{1}{{\sqrt {2\pi \alpha } }}e^{\pm ik_Fx}e^{i( \mp \phi  + \theta )}\; ,
\label{fermions_def}
\end{equation}
in terms of which $H_-$ becomes a free Hamiltonian. Note that the
``Fermi momentum" $k_F$ is dictated by $q$, which can be viewed as
a vortex filling factor in the line-junction area. Quite
interestingly, this implies that near a SIT, it is natural to
adapt a duel representation of this system in terms of {\it
fermionic} vortex fields. We note that the vortex--matter in the
system is constrain by an effective periodic potential set by the
vortex-vortex interaction and the boundary conditions: the leads
connected to $x=\pm L/2$ (assumed to be macroscopic
superconductors), and the strip edges which induce an effective
"image charges" potential \cite{likharev}. Hence, the vortices
tend to form a pinned chain where core positions are separated by
a constant spacing $a=L/N$, in which $N={\mathcal I}[qL]$
(${\mathcal I}[z]$ the integer value of $z$) denotes the total
number of vortices \cite{glazman}. We hereon regard Eq.
(\ref{H_-bos}) as a continuum limit of a lattice model, where the
coordinate $x=na$ ($n$ integer). Consequently, $q$ is replaced by
the deviation of the vortex density from the closest commensurate
value:
\begin{equation}
q=\frac{w(B-B_N)}{\Phi_0}\; ,\quad B_N=NB_0\,,\quad B_0\equiv \frac{\Phi_0}{wL}\,;
\label{new_q}
\end{equation}
$N={\mathcal I}[B/B_0]$ so that $0\leq q\leq \frac{1}{2}$. The lattice constant $a$ also determines the short-distance cutoff $\alpha$ in Eq. (\ref{fermions_def}).

The fermionic representation of $H_-$ is given by
\begin{multline}
H_-  = \int dx \{  v_ -  [\psi _R^\dag  (x)( - i\partial _x )\psi _R (x) - \psi _L^\dag  (x)( - i\partial _x )\psi _L (x)] \\
- \mu_v [\psi _R^\dag  (x)\psi _R (x) + \psi _L^\dag  (x)\psi _L (x)] \\
-J [\psi _R^\dag (x)\psi _L  (x) + \psi _L^\dag (x)\psi _R  (x)] \\
+ V [\psi _R^\dag (x)\psi _L^\dag (x) + \psi _L  (x)\psi _R
(x)]\} \label{H_-_ferm}
\end{multline}
where $J=\pi\alpha g_J$, $V=\pi\alpha g_c$ and the vortex chemical
potential $\mu_v$ is determined by the band-structure of the
Fermions. For simplicity, we associate the commensurate case $q=0$
with a $1/2$--filled band at $\mu_v=0$; e.g.,
$\mu_v=\epsilon_0\cos[\pi(1/2-q)]$ with $\epsilon_0\sim v_-/a$. We
note that the particular dependence of $\mu_v$ on $q$ is not
crucial to demonstrate the qualitative features of the model:
however generally, it is monotonically increasing with
$\mu_v(q=0)=0$.

Following the analogous problem of spin-$1/2$ ladders \cite{SNT,Tsvelik}, it is useful to decompose the complex Fermions [Eq. (\ref{fermions_def})] in terms of the Majorana fields
\begin{equation}
\eta _{1\nu} = \frac{1}{{\sqrt 2}}\left(\psi_\nu+ \psi _\nu^\dag \right)\; ,\quad \eta _{2\nu} = \frac{1}{i{\sqrt 2}}\left(\psi_\nu- \psi _\nu^\dag \right)
\label{majo_def}
\end{equation}
($\nu=R,L$). Recasting Eq. (\ref{H_-_ferm}) in $k$-space and using the Fourier transformed fields $\eta_{j\nu,k}=\eta_{j\nu,-k}^\dagger$, we obtain
\begin{multline}
\label{H_-_majo}
H_ -   = \sum_k \Psi_k^\dagger \mathcal{H}_k \Psi_k\; ,\\
\mathcal{H}_k \equiv
\left( {\begin{array}{*{20}c}
   {v_-  k } & { i\Delta_u^{(0)}} & {i\mu_v } & 0  \\
   {-i\Delta_u^{(0)}} & { - v_-  k } & 0 & { - i\mu_v }  \\
   { - i\mu_v } & 0 & {v_-  k } & {-i\Delta_d^{(0)}}  \\
   0 & {i\mu_v } & {  i\Delta_d^{(0)}} & { -v_-  k }  \\
\end{array}} \right) \\
\Psi_k^\dagger \equiv
\left( {\begin{array}{*{20}c}
   {\eta_{1R,k} }  & , &
   {\eta_{2L,k} }  & , &
   {\eta_{2R,k} }  & , &
   {\eta_{1L,k} }
\end{array}} \right)\; ;
\end{multline}
here $\Delta_{u,d}^{(0)}=J\pm V$ denote the gaps in the excitation
spectrum for commensurate vortex filling ($q=\mu_v=0$), in which
case $\mathcal{H}_k$ decouples into two independent blocks. Since
$J,V$ are positive, the $u$ sector is higher in energy.

We now focus on the case of interest, where the system is assumed
to be in the SC phase but close to a SIT so that the Josephson
energy $J$ is slightly larger than $V$, and $\Delta_{d}^{(0)}\ll
\Delta_u^{(0)}$. In this case, the high energy sector $u$ can be
truncated, and the low-energy properties are governed by the
$d$-type Fermions. In particular, the gap $\Delta_{d}^{(0)}$ can
{\it change sign} upon tuning of $J$ below the critical value
$J_c=V$ where $\Delta_{d}^{(0)}=0$. Indeed, for $\mu_v=0$ each
species of free massive Fermion models described by
(\ref{H_-_majo}) can be independently mapped to an Ising chain in
a transverse field \cite{Ising,QPT2}. When finite vortex ``doping"
is introduced by tuning $B$ away from $B_N$ such that
$\mu_v\not=0$, the original $d$ and $u$ sectors mix.
However, the resulting long wave-length ($k\rightarrow 0$) theory can still be cast in terms of two decoupled $d$ and $u$ sectors, where the energy spectrum has the same form but with modified velocities and gaps.
In particular, the modified gaps for finite $\mu_v$ are given by
\begin{equation}
\Delta_{u,d}(\mu_v)=J\pm \tilde V\; ,\quad \tilde V\equiv
\sqrt{V^2+\mu_v^2}\; . \label{gaps}
\end{equation}
While $\Delta_u$ remains positive and large for arbitrary $\mu_v$, a quantum phase transition occurs at a critical value of $\mu_v$ [which can be traced back to a {\it sequence} of critical fields $B_c^{(N)}$ via $\mu_v(q)$ and Eq. (\ref{new_q})], where $\Delta_{d}$ changes sign. As $B\rightarrow B_c^{(N)}$, $|\Delta_d|\sim |B-B_c^{(N)}|$. Below we show that these Ising--like quantum critical points correspond to SIT.

To study transport properties of the system it is necessary to
include a scattering potential, generically induced by random,
uncorrelated impurities along the coupled wires. Without loss of
generality, we hence consider random impurities in wire $1$ by
including a linear coupling of $\rho_1(x)$ to a disorder potential
$V_D(x)$ in the Hamiltonian. The leading contribution to
dissipation  arises from the backscattering term of the form
\cite{book}
\begin{equation}
H_D=\int dx \zeta(x)\cos\{2\theta_1(x)\}
\label{dis}
\end{equation}
[see Eq. (\ref{rhototheta})], where we assume $\langle \zeta(x)\rangle=0$, $\langle \zeta(x) \zeta(x')\rangle =
D \delta(x-x')$ [with $\langle\,\rangle$ including disorder averaging].

For sufficiently small $D$, a perturbative treatment of $H_D$
(see, e.g., Chap. 7.2 in \cite{book}) yields the d.c. resistance
at finite $T$
\begin{multline}
R(T)\approx\mathcal{A}\,D\int_0^\infty dt\, t\Im m \{\chi (t)\}\;
,\\ \chi(t)\equiv\langle
\sin\{2\theta_1(t)\}\sin\{2\theta_1(0)\}\rangle_0\; ;\label{R_def}
\end{multline}
$\mathcal{A}$ is a numerical constant, and $\langle\,\rangle_0$
denotes an expectation value with respect to $H_0$ [Eq.
(\ref{Hpm})]. Using $\theta_1=\frac{1}{\sqrt{2}}(\theta_+ +
\theta_-)$, the correlation function $\chi(t)$ decouples into
\begin{eqnarray}
\chi (t)&=&\chi_{C+}(t)\chi_{C-}(t)+\chi_{S+}(t)\chi_{S-}(t)\; ;\nonumber\\
\chi_{C\pm}&\equiv &\langle \cos\{\sqrt{2}\theta_\pm(t)\}\cos\{\sqrt{2}\theta_\pm(0)\}\rangle_\pm \label{chi_R} \\
\chi_{S\pm}&\equiv &\langle \sin\{\sqrt{2}\theta_\pm(t)\}\sin\{\sqrt{2}\theta_\pm(0)\}\rangle_\pm  \nonumber
\end{eqnarray}
where $\langle\,\rangle_\pm$ are evaluated w.r.t. $H_\pm$. Since $H_+$ is a Luttinger Hamiltonian [see Eq. (\ref{Hpm})], we obtain \cite{book}
\begin{equation}
\chi_{C+}(t)=\chi_{S+}(t)=\lim_{\epsilon\rightarrow 0}\left(\frac{-(\pi\alpha T/v_+)}{\sinh\{\pi T\left(t-i\epsilon\right)\}}\right)^{\frac{1}{K_+}}.
\label{chi_+}
\end{equation}
In contrast, as discussed below, $\chi_{C-}(t)$, $\chi_{S-}(t)$ depend crucially on the parameters of (\ref{H_-_majo}), and in particular on the magnitude and {\it sign} of the masses $\Delta_{u,d}$.

To evaluate $\chi_{C-}$, $\chi_{S-}$ we first note that in terms of the field $\theta$ [Eq. (\ref{new_theta})], they correspond to correlation functions of  $\cos\theta$, $\sin\theta$, which lack a local representation in terms of Fermion fields. However, a convenient expression is available in terms of the two species of order ($\sigma_{u,d}$) and disorder ($\tilde\sigma_{u,d}$) Ising fields \cite{SNT,Ising}: for $\Delta_d>0$,
\begin{equation}
\cos\theta\sim \sigma_u\tilde\sigma_d\; ,\quad \sin\theta\sim
\tilde\sigma_u\sigma_d\; . \label{Ising_udp}
\end{equation}
For $\Delta_d<0$, the roles of $\sigma_d$, $\tilde\sigma_d$ are simply {\it interchanged}.
The correlators $\chi_{C-}$, $\chi_{S-}$ can therefore be expressed in terms of $C_\lambda(t)=\langle \sigma_\lambda(t)\sigma_\lambda(0)\rangle$, $\tilde C_\lambda(t)=\langle \tilde\sigma_\lambda(t)\tilde\sigma_\lambda(0)\rangle$ ($\lambda=u,d$), which have known analytic approximations in the semi--classical regime ($|\Delta_\lambda|\gg T$) \cite{SNT,sachdev,BMASR}:
\begin{equation}
C_\lambda(t)\sim |\Delta_\lambda|^{1/4}K_0(i|\Delta_\lambda|t) ,\quad \tilde C_\lambda(t)\sim |\Delta_\lambda|^{1/4}
\label{C_tildeC}
\end{equation}
[with $K_0(z)$ the modified Bessel function]. In the quantum critical regime ($|\Delta_d|\ll T$), $C_d(t)\sim \tilde C_d(t)\sim t^{-1/4}$.

Employing Eqs. (\ref{R_def})--(\ref{C_tildeC}), we derive expressions for the low--$T$ resistance $R(T,B)$ near commensurate fields $B_N$ [Eq. (\ref{new_q})] where $\Delta_d=\Delta_d^{(0)}$, and near $B_{N+\frac{1}{2}}\equiv(N+\frac{1}{2})B_0$, where $\Delta_d$ is maximally negative. Neglecting terms of order $e^{-\Delta_u/T}$, we obtain for $B\sim B_N$
\begin{eqnarray}
R(T,B)&\approx & R_s\sqrt{\frac{\Delta_d(B)}{T}}\left(\frac{\pi\alpha \Delta_d(B)}{v_+}\right)^{-\kappa(B)}
e^{-\Delta_d(B)/T},\nonumber \\
\kappa(B)&\equiv & 2-K_+^{-1}(B)\; ,\quad R_s\propto D\left(\Delta_d(B)\right)^{1/4}\; .
\label{R_SC}
\end{eqnarray}
Superimposed on a moderate increase with $B$ arising from
$\kappa(B)$ due to the suppression of $\rho_s$ [Eq.
(\ref{Kvg_c})], the exponential factor leads to a strong {\it
decrease} and $R\rightarrow 0$ at $T\rightarrow 0$ as long as
$\Delta_d(B)>0$ is finite. The disordered Ising phase is thus
identified as {\it superconducting}, suggesting that the fields
$\sigma_d$ physically represent phase-slips. In contrast, for
$B\sim B_{N+\frac{1}{2}}$ ($\Delta_d<0$) we find
\begin{equation}
R(T,B)\approx  R_i\left(\frac{\pi\alpha |\Delta_d(B)|}{v_-}\right)^{1/4}\left(\frac{\pi\alpha T}{v_+}\right)^{-\kappa(B)}
\label{R_I}
\end{equation}
($R_i\propto D$). Since $K_+^{-1}\lesssim \frac{1}{2}$, $\kappa(B)>0$ yielding $R\rightarrow\infty$ at low $T$, indicative of an {\it insulating} behavior. In fact, in this regime the perturbative calculation leading to Eq. (\ref{R_I}) is not valid in the $T\rightarrow 0$ limit, where localization takes over and $R(T)$ diverges exponentially \cite{book}.

The above analysis implies that the quantum critical points at $B_c^{(N)}$ (where $\Delta_d=0$) correspond to SC-I and I-SC transitions alternately (see Fig. 1). However, note that unlike the 2D SIT, $R(T)$ in their vicinity does not manifest a metallic behavior: the power--law correlations in the critical regime yield
\begin{equation}
R(T,B)\sim T^{\gamma(B)}\; ,\quad \gamma(B)\equiv \frac{1}{4}-\kappa(B)<0\; .
\label{R_c}
\end{equation}
This reflects a slightly moderated insulating behavior, an asymmetry that stems from the quasi--1D nature of the system. As a result, a sharply defined crossing point of isotherms does not exist, as clearly indicated in Fig. 1.

To summarize, we have shown that prominent features of the magnetoresistance oscillations in SC strip--like devices are captured by a toy--model for the quantum Josephson--vortices in a line--junction between parallel SC wires. When the wires are close to a SIT, this system can be described by a field theory of 1D free Fermions, implying the existence of a sequence of quantum critical point of the Ising--type. It should be pointed out that since the Fermions are massive, small deviations from our ideal choice of parameters such that the Fermions become interacting do not change the essential properties, as the interactions can be treated perturbatively. The low--$T$ behavior of the resistance $R(T,B)$ provides a transparent interpretation of the critical points as transitions from a SC (near integer vortex fillings) to insulator (near $\frac{1}{2}$-integer vortex fillings) or vice versa. However, its behavior in the vicinity of the transitions is distinct from the SIT in fully 2D SC films. In particular, it does {\it not} reflect the duality symmetry of the underlying model, and the insulating regimes are effectively widened.

 {\it Final note added:} During the preparation of the manuscript we became aware of an independent work \cite{PGR}, proposing an alternative theoretical model as interpretation to the data of Ref. \onlinecite{shahar}. Both models yield qualitatively similar magnetoresistance oscillations.

We thank P. Goldbart, D. Pekker, Gil Refael, A. Tsvelik  and especially D. Shahar for useful discussions. E. S. is grateful to the hospitality of the Aspen Center for Physics. This work was supported by the Ministry of Science and
Technology grant No. 3-5792, BSF grant No. 2008256 and ISF grant No. 599/10.

\end{document}